# Organic Light-Emitting Diode Beam Shaping: Pixel Design for Variable Angular Emission Profile Control

*Felix Fries\*, Markus Fröbel\*, Pen Yiao Ang\*, Simone Lenk\*, and Sebastian Reineke\**
\*Dresden Integrated Center for Applied Physics and Photonic Materials (IAPP), Technische Universität Dresden, 01187 Dresden, Germany

**Abstract**
*Organic light-emitting diodes (OLEDs) are the leading self-emitting pixel technology in current and future small and large area displays. Once integrated with a certain layer architecture into the backplane layout, their emission colour and angular distribution is set by the optical properties of the layered system. In this paper, we demonstrate a pixel design that allows for actively controlled variation of the angular emission profile of the individual vertical pixel. For this, a tandem device is developed that comprises two units optimized for different angular emission pattern. We constrained the system to operate in a narrow emission band to maintain monochromaticity of the individual pixel. We discuss this concept for a red phosphorescence-based OLED stack and give an outlook based on simulations for the other primary display colours green and blue. The tandem unit can be operated with only two electrodes making use of the AC/DC driving concept, where the outer electrodes are in direct connection. In this paper, we will discuss the potential, status, and technology challenges for this concept.*

## 1. Introduction

Nowadays, organic light-emitting diodes (OLEDs) can be found in various fields of lighting applications, be it illumination, large area displays like televisions or even in micro displays. The inherent properties of OLEDs make them the unique possibility for many purposes. Devices are able to provide features like ultra-thin and flexible design, transparency, and spectrally broad white-light emission with a high colour-rendering index. Very often the emission intensity of such light sources follows a Lambertian like behaviour, which refers to a cosine-like drop of intensity with increasing viewing angle. However, this doesn't meet many application needs, which is why it often gets reshaped using secondary optical elements. The use of so-called beam-shaping methods spreads over many fields from laser physics, LED illumination, to rear lights of cars. Within the OLED community, however, broad beam-shaping concepts have been missing for long. Some ideas were presented in the past, using either microlens-arrays [1], diffractive gratings [2], or abandoning the planar geometry of the light source [3]. Especially the first two approaches don't make any use of the fundamental attributes of OLEDs, and thus will always be in strong competition with other light sources like LEDs. More important still, they all loose the great advantages of OLED light sources as mentioned at the beginning of this section. Having applications like micro displays in mind, particularly bulky beam-shaping structures should be avoided. Only recently we presented an active beam-shaping method, which can exclusively be applied to OLEDs, using their inherent properties, and hence adding nicely to their long list of advantages over alternative light-sources [4]. At that time, we showed a proof-of-concept, discussed upcoming challenges and geometrical influences of such a concept. Importantly, we focused our work mainly on red OLEDs.

Here, we transfer this promising concept to other primary colours. Especially when talking about display applications, all three colours of the CIE tristimulus need to be realised. Therefore, we first want to revise the most important theoretical features which are key for understanding. On the basis of experimental data of a green beam-shaping device, we discuss in the second part effects, which are not that prominent in red devices. Subsequent simulations, however, prove that those issues can be addressed choosing adequate emitter materials. In the language of device design, a blue emission colour can be seen as the extrapolation of the change from red to green and will be considered only briefly at the end of this text.

## 2. Theoretical Basics and Device Architecture

In Fig. 1 the device architecture is shown. The basic structure is shown on the right hand side and represents an AC/DC OLED [4, 5].

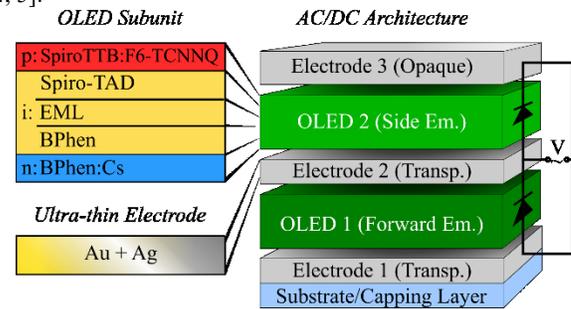

**Figure 1:** Left hand side: Each side unit is made up of a pin-OLED, comprising hole-transporting, electron-blocking, emission, hole-blocking, and electron transporting layers. The intermediate electrode is an ultra-thin wetting layer electrode. Right hand side: Two units are stacked on each other to build an AC/DC device.

This concept is characterised by two independent subunits (OLED 1 and OLED 2) which are stacked on top of each other. As the top electrode is connected to the bottom electrode, they remain as a common electrical counter pole to the transparent middle electrode. The latter is made of a thin gold wetting layer and a subsequent silver layer [6]. Depending on the polarity of the applied voltage only one of the subunits is in forward bias and thus, emitting light. Besides easy switching between the two devices, mixed emission can be achieved when applying an alternating voltage, providing a frequency higher than humans' perception. The flicker-fusion frequency can be assumed to be around 60 Hz [7]. Despite the fact that our samples were driven at 50 Hz, which corresponds to the main frequency in central Europe, we could not detect any flickering in the emission of the devices. Using pulse-width modulation, simple and continuous tuning between the pure emissions of each subunit is easily possible. This basic working principle is completely independent of the main emission direction of the final sample. It can either be bottom emitting, in which case electrode 1 would be indium tin oxide (ITO) and the emission happens through the glass substrate, or top emitting with the electrode 1 being a transparent Au/Ag layer with an additional organic capping layer for increased out-coupling of the light [8]. In the latter case, electrode 3 is directly located on the glass substrate. It is important to note that all our experiments and simulations show that independent of the emission direction, the sideward emitting unit should always be the one close to the thick opaque electrode 3. As hinted on the left hand side of Fig. 1, each of the two subunits is made up of a pin-OLED, referring to the electrical properties of the respective layers: a p-doped hole conducting layer, an n-doped electron conducting layer, and non-doped intrinsic layers for charge carrier confinement and light emission. The latter is also referred to as EML (emission layer). The material of choice depends on various parameter and will be discussed later in this work. The pin-layout is one of the key attributes to realize active beam-shaping, as the doped layers comprise both very high transparency and conductivity. Hence, varying their thickness can be used for flexible adjustments of the resonances in the micro-cavity [9].

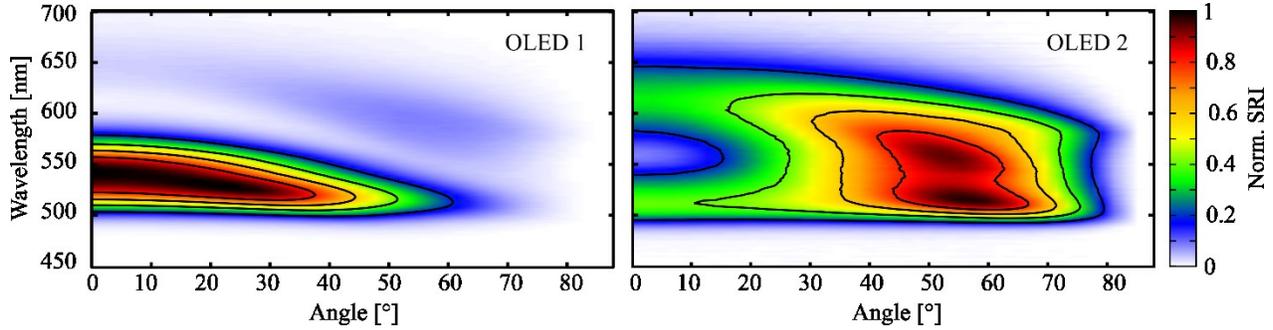

**Figure 2:** The experimentally measured normalized spectral radiant intensity of the two subunits of a green beam-shaping OLED. The subunit OLED 1 shows mainly forward emission, whereas OLED 2 has a clear intensity maximum at 56°. The spectra were individually normalized to their maximum intensity.

As shown in our previous work, the angular distribution of the emitted intensity depends strongly on the influence of the micro-cavity of the device, which are mainly the three following points: First, the overall thickness of the layers between the framing electrodes of the respective subunit determines the peak resonance wavelength. Second, the position of the EML within the cavity and its interaction with the electric field within the device influences both the efficiency and the angular behaviour of the emission. Third, following a Fabry-Pérot resonator behaviour, thicker electrodes lead to a spectrally more confined emission [4, 10].

Those influences can be summarized as the cavity mode (*CM*). By multiplication with the electroluminescence (EL) spectrum of the emitter, the shape of the totally out-coupled spectral radiant intensity (SRI) is obtained. The EL spectrum is normally assumed to resemble the photoluminescence (PL) spectrum, which is experimentally accessible. In total this gives the very simplified version of the SRI $I(\lambda, \theta)$, as given by Furno *et al.* [9]:

$$I(\lambda, \theta) \sim s_{EL}(\lambda) \cdot CM(\lambda, \theta)$$

Having this in mind, the strategy for the design of a beam-shaping OLED is optimizing the interplay of the *CM* with the emitter emission profile.

### 3. Experimental Findings

In our previous work, we demonstrate that due to the bending of the cavity modes towards shorter wavelengths with higher viewing angles, it is useful to use two different emitter materials in the subunits. OLED 1 which is the side emitting one, tends to be blue shifted compared to the emitter's peak PL intensity. To counteract to this effect an emitter is chosen for this subunit, which provides a red shifted PL spectrum as compared to the emitter in OLED 2. Experimental tests were done with the two well-known green emitters Ir(ppy)$_2$(acac) and Ir(ppy)$_3$. The shape of their PL spectra resemble each other quite well, whereas the peak wavelength is 523 nm and 507 nm, respectively. For detailed information on the materials please refer to the *Materials and Methods* section. The whole stack is built in top-emitting architecture, as here the cavity-mode is expected to be more confined and thus leads to a stronger side-emission.

The resulting SRI of each of the subunit is presented in Fig. 2. As expected, OLED 1 emits mainly into forward direction, which refers to low viewing angles. Its maximum is at 0° and 545 nm dropping to half intensity around 45°. On the other hand, OLED 2 shows its maximum of intensity at 56° and 516 nm. The two branches of the SRI at 0° arise from two contributions of the cavity mode. Integration of the SRI over the wavelength prove that the maximum intensity for OLED 2 is not at 0°, and thus results to be side emitting.

Even the projection of this radial symmetrical measurements onto a planar surface (see photo in Fig. 3a)) shows an intensity distribution with a maximum at angles larger than 0°. However, one big upcoming challenge becomes clear at this stage: as the contributions of different colours/wavelength vary differently as a function of the angle, a strong colour shift from the centre of the light spot towards the edges is visible. The brightness in each pixel is proportional to the grayvalue, which is the sum of the contributions of each primary colours (R, G, B). Figure 3b) shows the distribution of each colour channel and of the normalized grayvalue for a cross-section of the photo above. Even though the total brightness peaks around +/- 30°, the three colours differ in their behaviour. As the red channels peaks closer to 0°, than the green one, the conceived colour changes from reddish over strongly yellow to green. Of course, this is inacceptable for possible applications. A photo of the forward emission is not shown here, but as expected from the spectrum (Fig. 2), the intensity shows a point like behaviour with quite stable colour distribution.

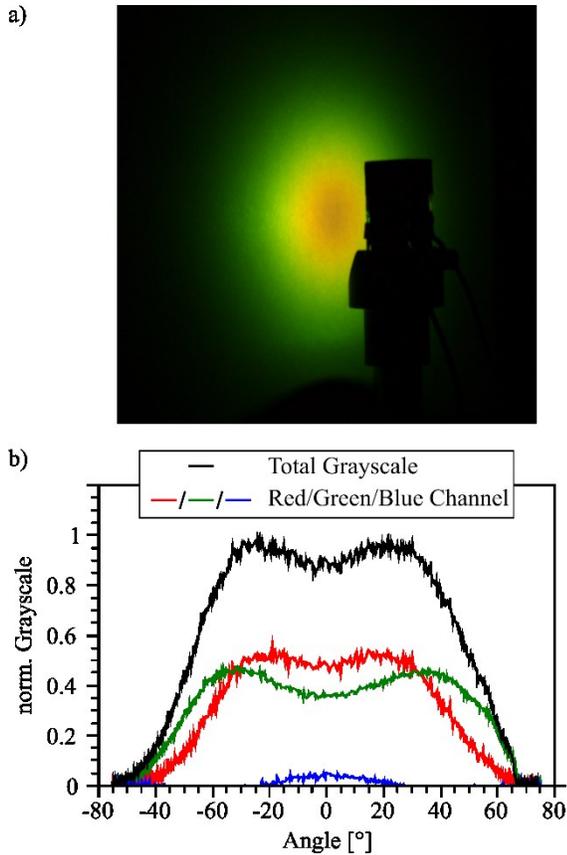

**Figure 3:** a) A photo taken of an illuminated screen shows a strong colour drift from a central yellow to green towards the edges. b) Even though the total emission shows maximum intensity at a non-zero viewing angle (black line), a separation into the primary colours shows their different angular behavior.

### 4. Simulation with Alternate Emitter Material

As this device is built in top-emission design, a quite strong cavity effect is already present in the sample. This means, to overcome the observed colour-shift, the best remaining possibility is to include emitting materials providing spectrally more narrow PL-spectra, as they are for example presented by G. Li *et al.* [11]. As those emitter materials were not at hand, we performed simulations assuming its emitting properties at this stage. The PL spectrum of the emitter PtN1N is shown in the inlet of Figure 4. It peaks around 500 nm and has a very narrow peak shape. This emitter was used in the side emitting OLED 2. As no anisotropy value is given in the publication, we assumed isotropic orientation (orientation factor a=0.33). As the forward emission is less critical, the emitter was not exchanged and remains Ir(ppy)$_2$(acac).

To make the results more comparable to the red beam-shaping OLEDs we've shown in our earlier work, and as it is the more common geometry, the following samples were simulated in bottom-emission architecture. Except for the emitting molecule, the used materials are unchanged to the stack shown above. The thickness of each transport layer was adjusted to obtain best spectra. The resulting emission as simulated when illuminating a flat screen is show in Figure 4. The shown angular range is +/-70° viewing angle. Two points are notable here. First, the two emission colours resemble each other quite well. Second, the colour drift in side emission could be reduced drastically, compared to the devices shown in the section before. Still there is a slight change from the centre to the maximum of intensity, but it is way less pronounced.

This simulation proves that narrow band emitter, which are already present in the community and for sure can expect further evaluation, can overcome the challenges which occur when transferring our beam-shaping concept to non-red devices.

### 5. Discussion

In this paper we took up the topic of active beam-shaping, we only presented recently for red OLEDs. It was shown that moving onto other emission colours like green, issues appear, which were not decisive before. Having a red emitter embedded in the device, dark red contributions are barely perceived by the human eye. Aiming for colours in the blue or green wavelength regime, long wavelength contributions, however, are easily visible. In the present case the green side emission was superimposed by a yellow central spot, which arises by red contributions in the spectrum at low viewing angles. Nevertheless, subsequent simulations show that narrow emission band emitter are able to reduce the parasitic contributions in the spectrum. Hence, the perceived image on a flat screen shows drastically reduced colour drift. Both the forward emission and the side emission shows nearly exactly the same green colour.

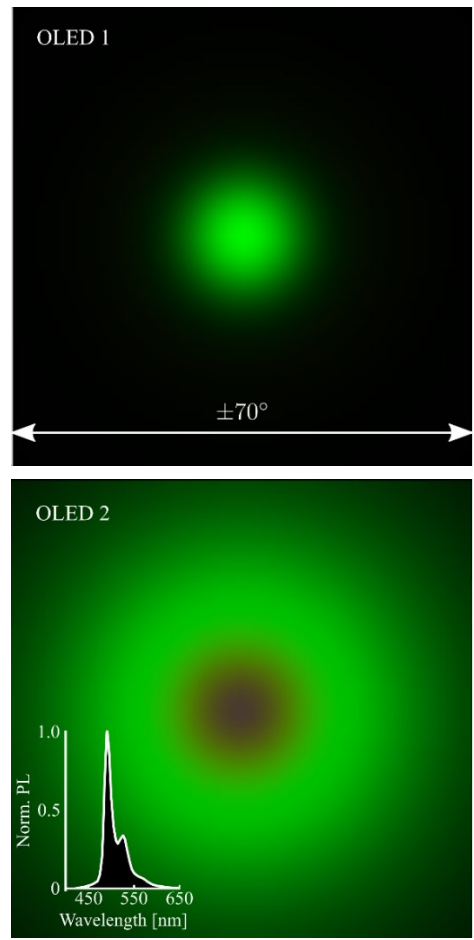

**Figure 4:** The simulated emission pattern using a narrow band emitter as can be found in the literature [10] (the PL spectrum is shown in the inset), allows for quite colour stable beam-shaping OLEDs.

The last missing piece towards white beam-shaping devices is a blue stack. However, we are confident, that the transfer from green to blue bares no more scientific difficulties. Still, from the production point of view the shorter the emission wavelength, the more precise the absolute thickness of each layer has to be realized.

Regarding display applications, not only the existence of all three primary colours is essential, but also the arrangement within the display's pixel. As one beam-shaping device already consists of two stacked units, which are having a cross correlation of their cavity modes, stacking three various beam-shaping devices would result in six stratified OLEDs. It is hard to see a way to realize this in a stable, reproducible, and cost-effective way. However, in small displays like smart-phone applications, the state of the art colour-mixing is achieved be placing the primary colours in different subpixels side by side. From the device point of view there is no reason why this should not be possible with beam-shaping devices. Having this in mind we are confident that the concept of active beam-shaping has high potential to be realised in future display applications.

## 6. Materials and Methods

The presented OLEDs were fabricated in a UHV chamber at a pressure around $10^{-6}$ to $10^{-7}$ mbar. The evaporation rates varied between 0.2 and 2 Å/s. Protection against water, oxygen and mechanical stress was achieved through glass to glass encapsulation.

The materials used are: N,N'-Di(naphthalen-1-yl)-N,N'-diphenyl-benzidine (NPB), silver (Ag), gold (Au), aluminium (Al), 4,7-diphenyl-1,10-phenanthroline (Bphen), caesium (Cs), 2,2',2"-(1,3,5-phenylen)tris(1-phenyl-1H-benzimida-zol) (TPBi), 4,4',4"-tris-(N-carbazolyl)triphenylamine (TCTA), Bis(2-phenylpyridine)iridium(III)acetylacetonate ($Ir(ppy)_2(acac)$), Tris(2-phenylpyridine)iridium(III) ($Ir(ppy)_3$), 2,2',7,7'-tetrakis-(N,N-diphenyl-amino)-9,9'-spirobifluorene (Spiro-TAD), 2,2',7,7'-Tetrakis-(N,N-dimethylphenylamino)-9,9'-spirobifluoren (Spiro-TTB), 2,2'-(perfluoronaphthalene-2,6-diylidene)di-malononitrile ($F_6$-TCNNQ).

The green top OLED follows the structure (from bottom to top, values in bracket refer to the thickness): glass-substrate / Al (40 nm) / Ag (40 nm) / Spiro-TTB doped with 4wt% of $F_6$-TCNNQ (105 nm) / Spiro-TAD (10 nm) / TCTA doped with 8wt% of $Ir(ppy)_3$ (8 nm) / TPBi doped with 8wt% of $Ir(ppy)_3$ (12 nm) / Bphen (10 nm) / Bphen doped with Cs (130 nm) [the doping is determined by a conductivity test to achieve a conductivity of $10^{-5}$ S/cm] / Au (2 nm) / Ag (7 nm) / Spiro-TTB doped with 4wt% of $F_6$-TCNNQ (40 nm) / Spiro-TAD (10 nm) / TCTA doped with 8wt% of $Ir(ppy)_2(acac)$ (8 nm) / TPBi doped with 8wt% of $Ir(ppy)_2(acac)$ (12 nm) / Bphen (10 nm) / Bphen doped with Cs (210 nm) / Au (2 nm) / Ag (7 nm) / NPB (60 nm).

The simulated bottom OLEDs show a very similar structure (again from bottom to top): glass-substrate / ITO (90 nm) / Spiro-TTB doped with 4wt% of $F_6$-TCNNQ (50 nm) / Spiro-TAD (10 nm) / TCTA doped with 8wt% of $Ir(ppy)_3$ (8 nm) / TPBi doped with 8wt% of $Ir(ppy)_3$ (12 nm) / Bphen (10 nm) / Bphen doped with Cs (170 nm) / Au (2 nm) / Ag (12 nm) / Spiro-TTB doped with 4wt% of $F_6$-TCNNQ (30 nm) / Spiro-TAD (10 nm) / TCTA doped with PtN1N (8 nm) / TPBi doped with PtN1N (12 nm) / Bphen (10 nm) / Bphen doped with Cs (250 nm) / Al (100 nm).

Spectral measurements were carried out in an in-house built spectro-goniometer sporting an USB-spectrometer (USB4000, Ocean Optics Inc.), pictures were taken using a Canon EOS D30 camera.

## 7. Acknowledgements

This project has received funding from the European Research Council (ERC) under the European Union's Horizon 2020 research and innovation programme (grant agreement No 679213).